# Performance of regression models as a function of experiment noise

Gang Li[1], Jan Zrimec[1], Boyang Ji[1], Jun Geng[1], Johan Larsbrink[1], Aleksej Zelezniak[1,2], Jens Nielsen[1,3,4], and Martin KM Engqvist[1*]

[1] Department of Biology and Biological Engineering, Chalmers University of Technology, SE-412 96 Gothenburg, Sweden
[2] Science for Life Laboratory, Tomtebodavägen 23a, SE-171 65, Stockholm, Sweden
[3] Novo Nordisk Foundation Center for Biosustainability, Technical University of Denmark, DK-2800 Kgs. Lyngby, Denmark
[4] BioInnovation Institute, Ole Måløes Vej 3, DK-2200 Copenhagen N, Denmark
* Corresponding author

E-mail: martin.engqvist@chalmers.se

## Abstract

A challenge in developing machine learning regression models is that it is difficult to know whether maximal performance has been reached on a particular dataset, or whether further model improvement is possible. In biology this problem is particularly pronounced as sample labels (response variables) are typically obtained through experiments and therefore have experiment noise associated with them. Such label noise puts a fundamental limit to the performance attainable by regression models. We address this challenge by deriving a theoretical upper bound for the coefficient of determination ($R^2$) for regression models. This theoretical upper bound depends only on the noise associated with the response variable in a dataset as well as its variance. The upper bound estimate was validated via Monte Carlo simulations and then used as a tool to bootstrap performance of regression models trained on biological datasets, including protein sequence data, transcriptomic data, and genomic data. Although we study biological datasets in this work, the new upper bound estimates will hold true for regression models from any research field or application area where response variables have associated noise.

# 1 Introduction

Machine learning (ML) has emerged as a powerful tool in biology and biological engineering [1–4] and is used to find hidden patterns in data (unsupervised learning) [5–8] as well as to discover complex relationships between sample features and labels (supervised learning) [9–14]. ML-based regression analysis, which is an example of supervised learning, is frequently applied in diverse biological fields including metabolic engineering [15,16], protein engineering [17,18], systems biology [19–23], environmental biology [10] and disease biology [24–26]. A key question when developing these supervised ML models is whether there is sufficient information in the available data to accurately predict sample labels. For a given dataset the performance of the best possible function for mapping input features to sample labels should thus be estimated to serve as a benchmark in ML model development. This level of performance is typically referred to as Bayes optimal error for classification problems [27]. In many classical ML problems - such as image classification, handwriting recognition and speech recognition - human-level performance at the task is very high and can therefore be used as a heuristic to estimate maximal performance [28,29]. However, for biological multi-dimensional data, human-level performance is low and is therefore not a good performance estimate. On the contrary, in biology one often seeks to train ML models for the explicit purpose of recognizing patterns and gaining insights that were obscured from the human intellect [30,31]. Therefore, without a clear performance benchmark against which to bootstrap biological regression models, it is difficult to know whether further model development is worth-while and when the performance limit has been reached.

When testing the performance of trained ML regression models, the discrepancy between predicted labels and observed labels in a test dataset is evaluated using metrics such as the mean squared error (MSE) and the coefficient of determination ($R^2$) [32]. Sample labels used in regression analysis of biological systems are typically real numbers obtained through measurements in a set of laboratory experiments. Such measurements inextricably have experimental noise and measurement error associated with them [33–35], thus affecting the quality of the sample labels. Because of such label noise a ML model with an MSE of 0 or $R^2$ of 1 (perfect prediction) cannot be achieved; there is an upper bound that cannot be surpassed. Methods to estimate this upper bound are underdeveloped, although some progress has been made recently [36,37]. Moreover, the resources invested into model development have diminishing returns on model performance as one approaches the upper bound. Knowing the best expected MSE or $R^2$ (i.e. the upper bounds) of a specific regression problem and dataset enables the discrepancy between current and potential model performance to be quantified, thus giving researchers a means to assess the cost-benefit trade-off of further model development.

In the present study, we mathematically derived a method to estimate upper bounds for the regression model performance metrics MSE and $R^2$ directly from the experimental noise associated with response variables in a dataset, independently of their predictors. Using Monte Carlo simulations, we show that this method is highly accurate and outperforms existing ones. Furthermore, we apply the method to real biological modeling problems and datasets, including protein sequence data, transcriptomics data and genomics data to demonstrate how the new upper bound estimates can inform model development.

## 2 Results

### 2.1 Estimating the theoretical upper bound of regression model performance

Starting from first principles, we mathematically derive a method to estimate upper bounds for model performance in terms of MSE and $R^2$. Given a set of samples with experimentally determined labels $\{y_{obs,i}\}$ and corresponding unknown real labels $\{y_i\}$, we assume a normally distributed experimental noise $\varepsilon_{y,i} \sim N(0, \sigma_{y,i})$: $y_{obs,i} = y_i + \varepsilon_{y,i}$ ($y_i \in R$), and that a complete set of features is known as $x_i \in R^k$ for each sample. This complete set of features can be used to calculate the real value of label $y_i$ with $y = f(x)$ for all samples. The performance of this real function $f(x)$ on the dataset $\{x_i, y_{obs,i}\}$ gives an upper bound for the expected performance of any ML model. The coefficient of determination ($R^2$) is a common metric to assess model performance and thus the $R^2$ of the model in the above argument is given by

$$R^2 = 1 - \frac{\sum_{i=1}^{m}(y_{obs,i} - \widehat{y}_{obs,i})^2}{\sum_{i=1}^{m}(y_{obs,i} - \overline{y}_{obs})^2} = 1 - \frac{\sum_{i=1}^{m}(y_{obs,i} - f(x_i))^2}{\sum_{i=1}^{m}(y_{obs,i} - \overline{y}_{obs})^2} \quad [1]$$

where $m$ is the number of samples. Although it is not possible to obtain an exact value from the above equation, since the real values $f(x_i)$ are unknown, we can instead obtain the expectation of $R^2$ (Supplementary Note 1), which is given by

$$\langle R^2 \rangle = 1 - \langle \frac{\sum_{i=1}^{m}(y_{obs,i} - f(x_i))^2}{\sum_{i=1}^{m}(y_{obs,i} - \overline{y}_{obs})^2} \rangle = 1 - \frac{m}{m-3} \frac{\overline{\sigma_y^2}}{\sigma_{obs}^2} \quad [2]$$

and in which $\overline{\sigma_y^2} = \frac{1}{m} \sum_{i=1}^{m} \sigma_{y,i}^2$. As the number of examples in machine learning is usually very large ($m \gg 1$), we can approximate the final equation for upper bound estimation as

$$\langle R^2 \rangle \approx 1 - \frac{\overline{\sigma_y^2}}{\sigma_{obs}^2} = \frac{\sigma_{obs}^2 - \overline{\sigma_y^2}}{\sigma_{obs}^2} \quad [3]$$

We refer to this upper bound estimate as $\langle R^2 \rangle_{LG}$ hereafter. It has a variance of $\frac{2m(m-2)}{(m-3)^2(m-5)} \frac{\overline{\sigma_y^4}}{\sigma_{obs}^4}$ (Supplementary Note 1). Similarly, the best mean squared error that one can expect is given as $\langle MSE \rangle = \overline{\sigma_y^2}$ with a variance of $\frac{2\overline{\sigma_y^4}}{m}$ (Supplementary Note 2).

The new upper bound estimate $\langle R^2 \rangle_{LG}$, which is the expectation of the $R^2$ of the best model that one can achieve via machine learning, solely depends on two properties of the dataset: (i) the true variance of the observed response values ($\sigma^2_{obs}$) and (ii) the average variance of experimental noise of all samples ($\overline{\sigma^2_y}$). In practice, $\sigma^2_{obs}$ and $\overline{\sigma^2_y}$ are unknown and have to be approximated from the dataset. $\sigma_{y,i}$ can be approximated with the standard error (SE) of $n$ replicates, which represent the standard error of the mean, and $\sigma^2_{obs}$ can be approximated as the variance of the target values (Figure 1). Since the resulting $\langle R^2 \rangle_{LG}$ is an expectation and relies on approximated values it does not strictly represent an upper bound for the $R^2$ of regression models and the real value may be slightly higher or lower. In this way the $\langle R^2 \rangle_{LG}$ estimate is analogous to using human-level performance to approximate upper bounds in image classification applications [38,39].

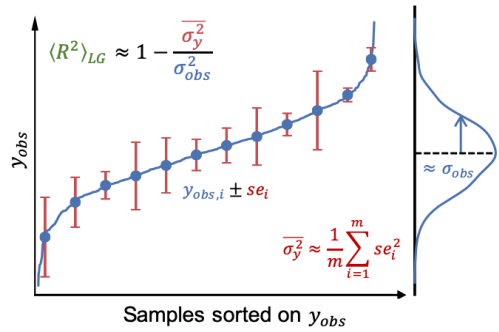

**Figure 1.** Schematic diagram depicting the estimation of the upper bound of model performance $\langle R^2 \rangle_{LG}$ based on experimental label noise. $\overline{\sigma^2_y}$ can be approximated from the standard errors (*se*) of samples in the dataset, and $\sigma^2_{obs}$ can be approximated as the variance of the target values. Data shown were randomly generated, $se_i$ denotes standard error of sample *i*.

## 2.2 $\langle R^2 \rangle_{LG}$ upper bound estimates outperform existing methods

Next, we benchmarked $\langle R^2 \rangle_{LG}$ upper bound estimates against existing methods. In two recent publications Fariselli and coworkers [36,37] proposed that given a set of experimentally measured values $y_{obs,i}$, the best possible model is $y = x$ in which $x$ are the values collected from another set of experiments conducted at identical conditions. Under this assumption, the expectation of the upper bound for MSE is $2\overline{\sigma^2_y}$ and $R^2$ is $\frac{\sigma^2_{DB} - \overline{\sigma^2_y}}{\sigma^2_{DB} + \overline{\sigma^2_y}}$, where $\overline{\sigma^2_y}$ is the average variance of all sample noise and $\sigma^2_{DB}$ is the variance of the real values (without noise). Since

$\sigma_{DB}^2 + \overline{\sigma_{\tilde{y}}^2} \approx \sigma_{obs}^2$, the upper bound for $\langle R^2 \rangle$ becomes $\frac{\sigma_{obs}^2 - 2\overline{\sigma_{\tilde{y}}^2}}{\sigma_{obs}^2}$, and we refer to this upper bound as $\langle R^2 \rangle_{FP}$ hereafter. In their publications, Fariselli and co-workers stated that no ML model could perform better than this upper bound. Comparing the equations for $\langle R^2 \rangle_{LG}$ and $\langle R^2 \rangle_{FP}$ it is clear that $\langle R^2 \rangle_{FP}$ estimates are lower than $\langle R^2 \rangle_{LG}$.

To directly compare $\langle R^2 \rangle_{FP}$ and $\langle R^2 \rangle_{LG}$, we performed Monte Carlo simulations. Briefly, a random dataset $\{x_i, y_{obs,i}\}$ was generated from a known real function $f(x)$ with added experimental noise $\sigma_{y,i}$. For this dataset $\langle R^2 \rangle_{FP}$ and $\langle R^2 \rangle_{LG}$ were calculated, and then the $R^2$ of a support vector machine regression model [40] trained on the data was calculated via a 2-fold cross validation approach ($R_{ML}^2$). This process was repeated for 1000 iterations. A linear (Figure 2A) and nonlinear real function (Figure 2B) were used in two separate simulations. Furthermore, to evaluate the effects of feature noise on regression model performance, we generated noise-free features ($\sigma_x^2 = 0.0$), as well as features with different levels of noise associated with them ($\sigma_x^2 = 0.2$, $\sigma_x^2 = 0.4$, $\sigma_x^2 = 0.6$, $\sigma_x^2 = 0.8$, $\sigma_x^2 = 1.0$). The simulations illustrated three key points. First, in both linear and nonlinear cases, $R_{ML}^2$ is always smaller than or close to $\langle R^2 \rangle_{LG}$, which confirms that $\langle R^2 \rangle_{LG}$ gives a good estimation of the model performance upper bound. Second, the simulations show that there are ML models with $R_{ML}^2$ higher than $\langle R^2 \rangle_{FP}$, which is contrary to the expacation if $\langle R^2 \rangle_{FP}$ is a true upper bound [36,37]. Third, as $\sigma_x^2$ increases, the ML model performance falls short of the $R^2$ upper bound, eventually falling below $\langle R^2 \rangle_{FP}$. This shows that $\langle R^2 \rangle_{LG}$ gives a more accurate estimation of the upper bound for the performance of ML models at any condition, including cases with or without noisy features. $\langle R^2 \rangle_{FP}$, is however useful as an estimate of the reproducibility of experiments, in accordance with the assumptions in the original papers [36,37].

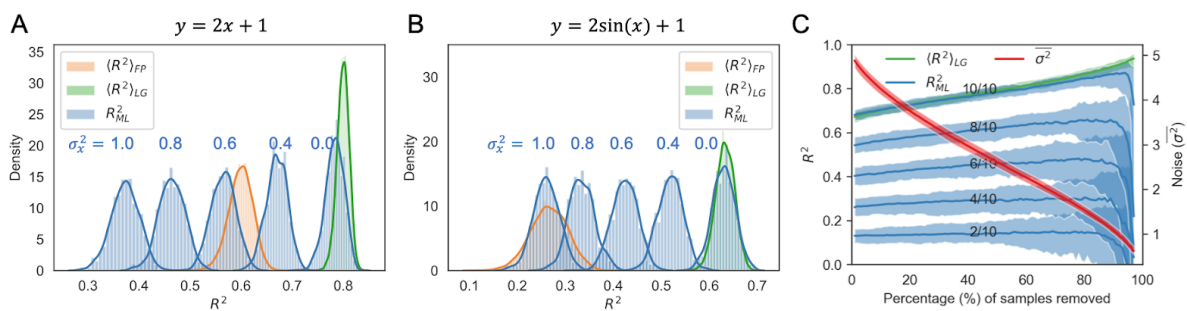

**Figure 2.** Monte Carlo simulation on the upper bound of $R^2$ assuming different levels of feature noise. $\langle R^2 \rangle_{FP}$ and $\langle R^2 \rangle_{LG}$ are expected upper bounds for $R^2$ with equations derived by Fariselli *et al* [36] and this study,

respectively. $R^2_{ML}$ is the $R^2$ obtained via a 2-fold cross-validation with a support vector machine. Two real functions were tested; (A) linear and (B) nonlinear. $\sigma^2_x$ is the variance of noise added to feature vector $x$. (C) Monte Carlo simulation on data cleaning via gradually removing the samples with the largest $\sigma_{y,i}$. n/10 indicate that n features out of a complete set of 10 features are used to train and validate the model. Noise values are given as the average variance of all samples ($\overline{\sigma^2}$).

In the above analysis idealized conditions were used in that all features were known. Conversely, in real-world machine learning applications, typically only an incomplete set of features is known. To more accurately simulate this real-world situation, we performed Monte Carlo simulations using incomplete feature sets. We also evaluated how ML model performance is affected by removal of the most noisy sample labels (Figure 2C). As anticipated, models trained with the full feature set (10/10) outperformed those trained with a subset of features, with the model containing all 10 features reaching the $\langle R^2 \rangle_{LG}$ (Figure 2C). Furthermore, model performance generally improved as noisy samples were removed. However, an interesting observation is that the degree to which the models improve upon removal of noisy samples depends on how many features were used to train them. For instance, if only a small fraction of relevant features were used (2/10 in Figure 2C), the removal of the most noisy samples did not improve model performance. In contrast, when a majority of the relevant features were known (8/10 and 10/10 in Figure 2C), the removal of noisy samples significantly improved the model performance in terms of $R^2$. These results indicate that when $R^2_{ML}$ is very far from the $\langle R^2 \rangle_{LG}$ upper bound, model performance can be most readily improved by obtaining additional or more relevant features, as opposed to performing data cleaning to reduce sample noise.

## 2.3 Enzyme $T_{opt}$: Using the theoretical upper bound to inform modeling

We next explored the applicability of $\langle R^2 \rangle_{LG}$ to inform ML model development on real-world data. The goal was to obtain models to accurately predict enzyme optimal catalytic temperatures ($T_{opt}$) using features extracted from their protein primary structures. A dataset comprising the $T_{opt}$ of 5,343 individual enzymes was collected from the BRENDA[41] database. Using enzymes for which $T_{opt}$ values had been measured in multiple experiments the experimental noise $\overline{\sigma^2_y}$ was estimated as $(7.84\,°C)^2$ and $\sigma^2_{obs}$ was $(16.32\,°C)^2$ in this dataset. Given these values for $\overline{\sigma^2_y}$ and $\sigma^2_{obs}$ the corresponding $\langle R^2 \rangle_{LG}$ upper bound was 0.77.

To provide features for ML model training we applied two established feature extraction methods on the 5,343 protein primary structures, one based on domain knowledge (iFeature [42], 5,494 features) and the other based on embeddings obtained from unsupervised deep learning (UniRep [43], 5,700 features). The ability of six different types of regression algorithms to predict enzyme $T_{opt}$ using those two feature sets was tested, and the result represented with the average $R^2$ score of 5-fold cross-validation. As can be seen in Figure 3A, models trained with features derived from iFeature achieved over 3-fold higher $R^2$ - score than those trained using UniRep features ($R^2$ was 0.42 and 0.13, respectively). However, even the best $R^2$ achieved (0.42) was only 55% of the estimated $\langle R^2 \rangle_{LG}$ upper bound (0.77) for this dataset, indicating that the model could be further improved. Such improvement might be achieved through either feature engineering or noise reduction, as seen in the Monte Carlo simulations (Figure 2C).

First, we performed feature engineering by including the optimal growth temperature (OGT) of the organism from which the enzyme was derived as an additional feature into the iFeature and UniRep feature sets. This came at the expense of having to omit 55% of the samples in the dataset (down to 2,917 enzymes) as they did not have known OGT values. Models trained with the new iFeature and UniRep feature sets, were improved by 33% and 384%, respectively (Figure 3A). The best resulting $R^2$ (0.56) achieved 71% of the estimated $\langle R^2 \rangle_{LG}$ (0.79). These results are consistent with our previous work [10], where it was shown that prediction of enzyme $T_{opt}$ was significantly improved when including OGT as a feature. In contrast to traditional ML models, deep neural networks automatically learn appropriate features from data [21,44]. To test whether a neural network could discover additional features in the enzyme primary structures, and provide better predictive models we trained a deep convolutional network (Figure S3) on the dataset and with/without OGT as an additional feature. This did not, however, lead to models that outperform the best classical ML models trained on iFeatures (Figure 3A).

Next, we considered reducing the noise ($\overline{\sigma_y^2}$) in $T_{opt}$ values as a means to further improve model performance. Using information from the "comment" field associated with each enzyme in the BRENDA database we removed values that were deemed less likely to represent true catalytic optima (see Methods for details). This process dramatically reduced the number of samples in the dataset and left only 1,902 enzymes from the initial 5,343, of which 1,232 were with known OGT. However, the experimental noise $\overline{\sigma_y^2}$ was reduced from

$(7.84\,°C)^2$ to $(7.22\,°C)^2$ and the calculated $\langle R^2 \rangle_{LG}$ increased from 0.79 to 0.86. Accordingly, the best model obtained with this dataset achieved an improved $R^2$ of 0.61, which again was around 71% of $\langle R^2 \rangle_{LG}$. On this dataset the deep network had a lower $R^2$ score than on the other two datasets (Figure 3A), this is in accordance with the expectation that large training datasets are required for optimal deep learning performance [44].

Finally, we performed an in-depth analysis to explore how different features of the iFeature set contributed to predictive accuracy. The 5,494 features from iFeature were broken up into 20 sub-feature sets according to their types, and their predictive power was evaluated using five different regression models. From this analysis we draw two main conclusions. First, we found that for each of these sub-feature sets reducing the noise in $T_{opt}$ only improved model performance when OGT was included as an additional feature (Figures 3B and 3C). This was consistent with results revealed in Monte Carlo simulations (Figure 2C), showing that noise reduction is only beneficial with more complete feature sets. Second, models trained using amino acid composition (20 features) showed the same predictive performance as the whole iFeature set, both with and without OGT as an extra feature, as well as before and after data cleaning (Figures S1 and S2). This showed that, compared to the amino acid composition, the 5,454 additional features derived from the protein sequence did not carry additional information for predicting enzyme $T_{opt}$. Future improvement of $T_{opt}$ prediction therefore necessitates that more relevant features are engineered, for instance ones extracted from protein 3D structures.

As a note, we used the improved model (a random forest trained on amino acid composition and OGT) to update $T_{opt}$ annotation of BRENDA enzymes in the Tome package [10] and also extended it to carbohydrate-active (CAZy) enzymes[45] (Figure S4) (https://github.com/EngqvistLab/Tome).

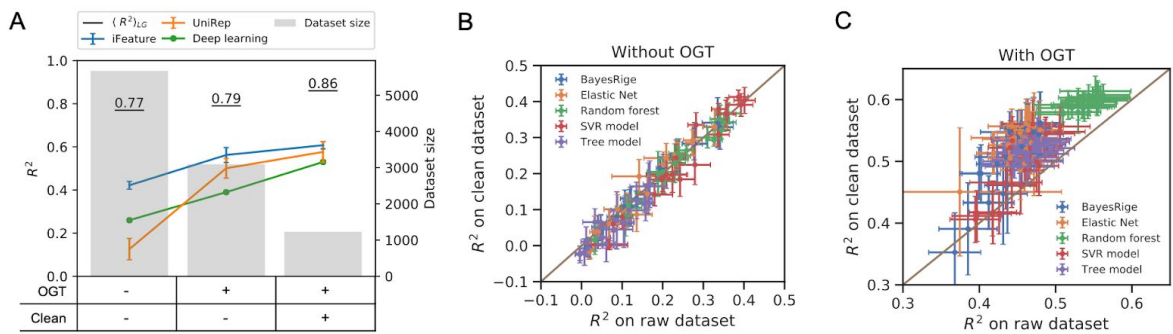

**Figure 3.** Development of the machine learning models for the prediction of enzyme optimal temperature ($T_{opt}$). (A) Performance of classical models using iFeatures [42] and UniRep encoding [43] feature sets as well as a deep neural networks with one-hot encoded protein sequence as input. (B, C) Comparison of model performance on raw and clean dataset (B) with; and (C) without OGT as one of the features. The features calculated by iFeature were grouped into 20 sub-feature sets as described in the Method section. Error bars show the standard deviation of $R^2$ scores obtained in 5-fold cross validation.

## 2.4 Transcriptomics: Amount of experimental noise affects estimates of $\langle R^2 \rangle_{LG}$ and model performance

We next explored how the amount of experimental noise in the response variables can affect the $\langle R^2 \rangle_{LG}$ and model performance. For this, we chose the problem of prediction of intrinsic gene expression levels in *Saccharomyces cerevisiae*, since thousands of transcriptomics RNAseq experiments across multiple conditions are available for this species [46]. For a given gene, the intrinsic expression level was defined as the average expression level across the different experiments and conditions [21]. The noise level could then be adjusted by increasing or decreasing the number of sampled data points (i.e. replicates) (Figure 4 inset), where the corresponding standard deviation was used to quantify the amount of noise present within the intrinsic gene expression levels (Methods). Apart from estimating the corresponding $\langle R^2 \rangle_{LG}$ upper bounds, the achievable predictive performance was explored by building linear regression models using DNA sequence features (codon usage) [21,47] as input variables and mRNA levels as the target variable (Methods).

We observed a strong effect of the amount of experimental noise on the theoretical upper bound, especially with a smaller number of data replicates (Figures 4 and S5). For example, the average $\langle R^2 \rangle_{LG}$ fell to 0.9 and below with 5 replicates or less, whereas with a larger number of replicates - e.g. 100 or above, the average $\langle R^2 \rangle_{LG}$ surpassed 0.99. Similarly, the variability of the $\langle R^2 \rangle_{LG}$ upper bound also markedly decreased with an increasing number of replicates, as with 3 replicates the standard deviation was almost 0.1 and decreased over 100-fold with 100 replicates or above (Figure 4: below $10^{-3}$). Therefore, with an insufficient amount of replicates, apart from a lower confidence in the estimated upper bound, the variability of the data were found to also drastically affect the predictive performance and accuracy of the models. For example, the average test $R^2$ obtained with the linear models increased from 0.62 with 3 replicates to above 0.7 when using 50 replicates or more. Similarly, the standard deviation of the test $R^2$ decreased 6-fold from 0.06 with 3 replicates to below $10^{-2}$ with 100 replicates, and fell below $10^{-3}$ only after the amount of replicates surpassed 1000. Therefore, for accurate condition-specific or cross-condition modeling, the number of replicates of at least 100, with most reliable results above 1000, should be used (Figure 4 and S5). Such dataset sizes are nowadays highly feasible, at least in the case of genomics, transcriptomics and proteomics data, as resources that comprise thousands of experiments are readily available for each model organism [46,48,49].

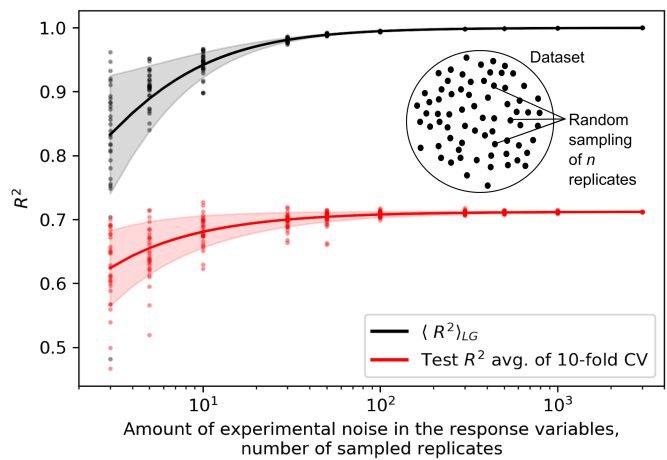

**Figure 4.** Analysis of the effect of experimental noise in the response variables on the $\langle R^2 \rangle_{LG}$ upper bound estimates (black) and predictive performance of ML models (red) with the case of a large yeast multi-experiment transcriptomics dataset [46]. The noise level was varied by adjusting the number of data replicates with random sampling (inset figure). Lines and shaded areas depict means and standard deviations of the 30 measurements per each *n* replicates, depicted as points.

## 2.5 From Genotype to Phenotype: $\langle R^2 \rangle_{LG}$ is applicable even in cases when experimental noise is unknown

For some datasets it may not be feasible to estimate the experimental noise, for instance if the values for replicates in an experiment are not available. We therefore show how $\langle R^2 \rangle_{LG}$ can be used to define the predictive potential of biological regression analysis in the absence of direct experimental noise estimates. Since $\langle R^2 \rangle_{LG}$ is an upper bound estimate $R^2_{ML} \leq \langle R^2 \rangle_{LG}$ holds true. From this we obtain that $\overline{\sigma_y^2} \leq (1 - R^2_{ML}) \times \sigma^2_{obs}$. If there are multiple datasets with the same level of (unknown) experimental noise are available the inequality holds for all datasets, meaning that $\overline{\sigma_y^2} \leq min(\{(1 - R^2_{ML,i}) \times \sigma^2_{obs,i} | i = 1,..,s\})$, in which $s$ is the number of the datasets. In this way it is possible to estimate the maximal level of the experimental noise based on the machine learning results, and then further use it to obtain the minimal value of $\langle R^2 \rangle_{LG}$ (referred to as $\langle R^2 \rangle_{LG,min}$). The in this special case $\langle R^2 \rangle_{LG}$ could be any value between $\langle R^2 \rangle_{LG,min}$ and 1.0. $\langle R^2 \rangle_{LG,min}$ would be useful when $\langle R^2 \rangle_{LG,min}$ approaches 1.0 and one can use it to check if there is still room to further improve $R^2_{ML}$ for some datasets.

We applied this idea for a case predicting yeast phenotypes directly from genomes (Figure 5A), which is a well-studied but very challenging task [50–54]. The dataset used in this section was taken from Peter J et al [49], in which the growth profiles of 971 sequenced *S. cerevisiae* isolates under 35 stress conditions had been measured. In the original paper, the experimental noise was not reported, nor were data for all replicates in the experiment provided. To analyze these data we made use of the *S. cerevisiae* pan-genome constructed in a previous study [48]. This pan-genome included all protein-coding genes across the 971 isolates with measured phenotypes. For ML analysis we chose to represent this pan-genome as either a gene presence/absence table (P/A, Figure 5B), or copy number variation table (CNV) which contains additional information to P/A (Figure 5C). The ML predictive performance of the 35 quantitative traits (experimental stress conditions) with P/A and CNV was tested with a random forest regressor. P/A and CNV showed a similar predictive power and could explain at most 30% of the variance (Figure 5D: $R^2$ was ~0.3) for some traits like the growth profile under the YPD medium enriched with 40 mM of caffeine (YPDCAFEIN40).

With $R^2_{ML}$s for these 35 datasets, the maximal experimental noise were estimated as $\overline{\sigma^2_y} \leq 0.076^2$, based on which we could finally estimate the $\langle R^2 \rangle_{LG,min} \approx 1 - \frac{0.076^2}{\sigma^2_{obs,i}}$ for each condition (Figure 5C). Despite that for a small number of traits (e.g. YPDNACL15M), the estimated minimal $\langle R^2 \rangle_{LG}$ was not useful as the real $\langle R^2 \rangle_{LG}$ could be any value between 0.05 and 1.0 (Figure 5D), in most cases it was higher than the current predictive performance of our models (e.g. > 0.97 with YPDCUSO410MM). Therefore, for most of the conditions, the estimated upper bounds showed great potential for further improvement of model performance (Figure 5D). For instance, the simple and incomplete description of the genome features used here (P/A or CNV of protein-coding genes) could in the future be upgraded with more comprehensive genome encodings that include SNPs [55] as well as chromosomal rearrangements [56].

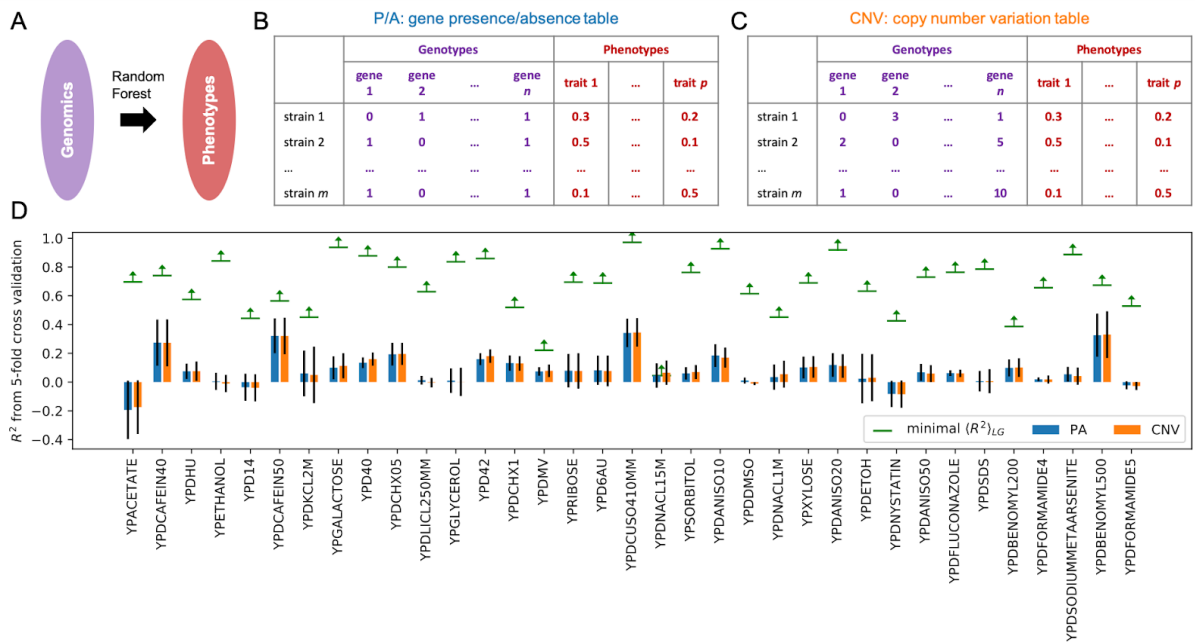

**Figure 5.** Prediction of 34 quantitative traits of *S. cerevisiae* from its pan-genome composition. (A) A random forest model applied to predict yeast phenotypes from genomics features. Genomes are represented as (B) gene presence/absence table and (C) copy number variance table in the pangenome [48]. (D) Obtained $R^2$ score for 35 different phenotypes. Experimental trait values were taken from [49]. Detailed label description can be found in Table S2 of [49]. Error bars show the standard deviation of $R^2$ scores obtained in 5-fold cross validation.

# 3 Discussion

In the present study, we established a framework to estimate an upper bound for expected ML model performance on regression problems. This addresses an important need in the ML field as human performance on multi-dimensional data is poor and cannot be used to bootstrap regression model performance [30,31], an approach that is common when developing ML systems for image analysis [28,29]. The coefficient of determination upper bound (model performance) for regression analysis is: $\langle R^2 \rangle_{LG} \approx 1 - \frac{\overline{\sigma_y^2}}{\sigma_{obs}^2}$, and depends only on the experimental noise $\overline{\sigma_y^2}$ and the variance of observed labels $\sigma_{obs}^2$ (Figure 1). This $\langle R^2 \rangle_{LG}$ is valid under the following assumptions: (1) observed label values $\{y_{obs,i}\}$ are normally distributed; (2) each $y_{obs,i}$ has a normally distributed experimental noise with zero mean. This upper bound was confirmed using Monte-Carlo simulations (Figure 2A) and was also shown to outperform existing measures [36] (Figure 2B). Calculating the $\langle R^2 \rangle_{LG}$ upper bound estimate for experimental data yields a more realistic modeling objective than naively assuming that an $R^2$ of 1.0 is possible. For instance, in the prediction of enzyme optimal temperature, $\langle R^2 \rangle_{LG}$ was estimated at 0.86 for a specific dataset (Figure 3A). Therefore, one should not expect to obtain ML models with $R^2$ scores above this value.

If the estimated $\langle R^2 \rangle_{LG}$ upper bound for a specific problem and dataset is low (label values are noisy compared to the label variance, $\overline{\sigma_y^2}$ is close to $\sigma_{obs}^2$), it may not be worthwhile to attempt modeling at all. An example of this is the prediction of melting temperatures of human proteins (Table S1) using the dataset from Leuenberger *et al* [57]. The sample labels for human proteins in this dataset has a large level of noise ($\overline{\sigma_y^2}$, 5.49²) compared to the label variance ($\sigma_{obs}^2$, 6.57²) and the calculated $\langle R^2 \rangle_{LG}$ was therefore correspondingly low at approximately 0.30. Even if a ML model with upper bound performance could be developed for these data, it would have little predictive value. In contrast, for three other, non-human, organisms in the Leuenberger dataset the calculated $\langle R^2 \rangle_{LG}$ was above 0.90, indicating that the development of predictive ML models may be worthwhile (Table S1). For data that are inherently noisy, such as those obtained from transcriptomics, the $\langle R^2 \rangle_{LG}$ upper bound, as well as the overall ML performance, can both be improved by increasing the number of experimental replicates used in the computational analysis (Figure 4).

If, on the other hand, the estimated upper bound $\langle R^2 \rangle_{LG}$ is high, the challenge of obtaining a model which achieves it still remains. Achieving upper bound performance is only possible when a complete set of noise-free features relevant for the predicted labels are used for model training and prediction (Figure 2A-C). When noisy features are used, the performance attainable by ML algorithms will be lower than $\langle R^2 \rangle_{LG}$ (Figures 2A and 2B), though the extent of this cannot be mathematically quantified as it would require knowledge of the real function $f(x)$. Furthermore, obtaining or engineering complete feature sets for biological samples is challenging as the labels are usually affected by a multitude of unknown factors. Thus, for classical ML models that rely on human feature engineering, models with test $R^2$ close to their upper bound are not easily achieved in practice (Figures 3A, 4 and 5D). In contrast, deep learning models can extract features directly from data [58], potentially resolving the issue of incomplete feature sets. However, deep learning requires a very large number of samples [59–61], something that is often difficult to obtain in biological studies. Using few samples when training deep models leads to suboptimal performance. An example of this can be seen in Figure 3A, where deep learning models did not perform on par with classical ones for enzyme $T_{opt}$ prediction when trained on a small dataset. When the sample number is limiting one can consider applying either data augmentation methods [62–64] or transfer learning, which uses part/whole of a pre-trained model on a large dataset, and then re-purposes it to another similar problem with a small amount of training samples [11,43,65].

To conclude, our method for estimating upper bounds for model performance should aid researchers from diverse fields in developing ML regression models that reach their maximum potential.

# 4 Methods

## 4.1 Brenda dataset

**Raw dataset:** We firstly collected $T_{opt}$ of 5,675 enzymes with known protein sequences from BRENDA [41]. Of these 3,096 enzymes were successfully mapped to a microbial optimal growth temperature database [66].

**Cleaned dataset:** To obtain a clean dataset with less noise we carried out several steps: (1) Enzymes for which the $T_{opt}$ entry contained "assay at" in the BRENDA "comments" field were removed from the raw dataset. (2) If a subset of all enzymes from a specific organism had the same EC number and exactly the same $T_{opt}$, then these were removed. This was done to address an issue with non-perfect matching between experimental data from the literature and Uniprot identifiers (186 enzymes). (3) Enzymes with multiple $T_{opt}$ values having standard deviations greater than 5 were removed (96 enzymes). After these steps 1,902 enzymes remained in the cleaned dataset, of which 1,232 were with known OGT. In both raw and cleaned datasets, any sequences shorter than 30 residues or contain letters that are not in 20 standard amino acids were discarded and for enzymes still with multiple $T_{opt}$ values the average value was used.

**Estimation of label noise:** For enzymes with multiple $T_{opt}$ values in BRENDA, the variance for each enzyme was calculated. Subsequently the average variance for all those enzymes was calculated and used as the estimation of experimental noise $\overline{\sigma_y^2}$ of the dataset. For the cleaned dataset the label noise was estimated at step (2) in the paragraph above, before samples with high standard deviation were removed.

## 4.2 Transcriptomics data

Genomic data including open reading frame (ORF) boundaries of *Saccharomyces cerevisiae* C288 was obtained from the Saccharomyces Genome Database (https://www.yeastgenome.org/) [67,68] and published data [69,70]. Coding regions were extracted based on ORF boundaries and codon frequencies were normalized to probabilities. Processed raw RNA sequencing Star counts were obtained from the Digital Expression Explorer V2 database (http://dee2.io/index.html) [46] and filtered for experiments that passed quality control. Raw mRNA data were transformed to transcripts per million (TPM) counts [71] and genes with zero mRNA output (TPM < 5) were removed. Prior to modeling, the mRNA counts were Box-Cox transformed [72].

### 4.3 Genomics data

The gene presence/absence (P/A) encoding of *S. cerevisiae* pan genome were obtained from Li G *et al*. [48]. The 35 quantitative traits were obtained from Peter J *et al*.[49].

### 4.4 Monte Carlo Simulations on expected $R^2$ score

Given the true functions between features and labels $f(x)$:

(1) Randomly generate 1,000 samples from $N(0, 1)$ as $x$. Then true values are $y = f(x)$;

(2) Randomly generate a noise vector $\varepsilon_y$. Each $\varepsilon_{y,i}$ is randomly sampled from $N(0, \sigma_{y,i}^2)$, where $\sigma_{y,i}^2$ is randomly sampled from $\chi^2(1)$;

(3) $y_{obs} = y + \varepsilon_y$;

(4) Add noise to $x_{obs} = x + \varepsilon_x$, in which $\varepsilon_x$ is sampled from a normal distribution with zero-mean and variance of $\sigma_x^2$ (varying from 0 to 1);

(5) Calculate $R_{ML}^2$ by performing a 2-fold cross-validation on dataset $\{x_{obs,i}, y_{obs,i}\}$ with support vector machine regression model (another inner 2-fold cross-validation for hyper-parameter optimization);

(6) Calculate upper bound for $R^2$ with $\langle R^2 \rangle_{LG} = \frac{\sigma_{obs}^2 - \overline{\sigma_y^2}}{\sigma_{obs}^2}$ and $\langle R^2 \rangle_{BS} = \frac{\sigma_{obs}^2 - 2\overline{\sigma_y^2}}{\sigma_{obs}^2}$, where $\sigma_{obs}^2$ is the variance of $y_{obs}$ and $\overline{\sigma_y^2}$ is the average value of randomly generated $\sigma_{y,i}^2$.

(7) Repeat above (1)-(6) for 1,000 times.

A linear function $f(x) = 2x + 1$ and a nonlinear function $f(x) = 2sin(x) + 1$ were tested respectively.

## 4.5 Monte Carlo simulations on data cleaning

Define a linear function $f(x) = \sum_{i=1}^{10} x_i$ as the true function to map 10 features to a target $y$. Each feature follows a standard normal distribution.

(1) Randomly generate feature of 1000 samples as $X$. Calculate real target values $y$;

(2) Randomly generate a noise vector $\varepsilon_y$. Each $\varepsilon_{y,i}$ is randomly sampled from $N(0, \sigma_{y,i}^2)$, where $\sigma_{y,i}^2$ is randomly sampled from $\chi^2(5)$;

(3) Calculate observed target values via $y_{obs} = y + \varepsilon_y$, and resulted a dataset $\{X, y_{obs}\}$;

(4) Assume we only know the first $n$ features ($n = 2, 4, 6, 8, 10$), Sort all samples based on $\sigma_{y,i}^2$ values, gradually remove the samples with the highest $\sigma_{y,i}^2$ values, calculate $R^2$ score of a linear function via a 2-fold cross validation on such a dataset with only subset of features.

(5) Repeat step 1 through 4 for a total of 100 times.

## 4.6 Feature extraction for enzymes in $T_{opt}$ dataset

A total of 20 different sets of protein features were extracted with iFeature [42] using default settings: amino acid composition (AAC, 20 features), dipeptide composition (DPC, 400), composition of k-spaced amino acid pairs (CKSAAP, 2400), dipeptide deviation from expected mean (DDE, 400), grouped amino acid composition (GAAC, 5), composition of k-spaced amino acid group pairs (CKSAAGP, 150), grouped dipeptide composition (GDPC, 25), grouped tripeptide composition (GTPC, 125), Moran autocorrelation (Moran, 240), Geary autocorrelation (Geary, 240), normalized Moreau-Broto (NMBroto, 240), composition-transition-distribution (CTDC, 39; CTDT, 39; CTDD, 195), conjoint triad (CTriad, 343), conjoint k-spaced triad (KSCTriad, 343), pseudo-amino acid composition (PAAC, 50), amphiphilic PAAC (APAAC, 80), sequence-order-coupling number (SOCNumber, 60) and quasi-sequence-order descriptors (QSOrder, 100). In total we obtained 5494 features from iFeature. Furthermore, we additionally obtained features in the form of sequence embedding representations encoded by a deep learning model UniRep [43], which is a Multiplicative Long-Short-Term-Memory (mLSTM) Recurrent Neural Networks (RNNs) that was trained on the UniRef50 dataset [73]. A total of $1900 \times 3$ features were extracted for each protein sequence using UniRep.

## 4.7 Supervised classical ML methods

Input features were firstly scaled to a standard normal distribution by $x_{N,i} = \frac{x_i - u_i}{\sigma_i}$, where $x_i$ is the values of feature i of all samples, $u_i$ and $\sigma_i$ are the mean and standard deviation of $x_i$, respectively. Two linear regression algorithms BayesianRidge and Elastic Net as well as three non-nonlinear algorithms Decision Tree, Random Forest and Support Vector Machine [27] were evaluated on each feature set (iFeatures and UniRep). The evaluation was conducted via a nested cross-validation approach: an inner 3-fold cross validation was used for the hyper-parameter optimization via a grid-search strategy and an outer 5-fold cross-validation was used to estimate the model performance (see Table S2C for hyper-parameter values). With the transcriptomics data, linear regression was the only algorithm used, as it was previously found to outperform all other algorithms with a similar dataset [21]. For genomics datasets, only the random forest regression was tested. All ML analysis was performed with scikit-learn (v0.20.3)[74] using default settings and Python v3.6.7.

## 4.8 Supervised deep ML methods

To test the performance of a deep neural networks on the prediction of enzyme $T_{opt}$, architectures were tested that comprised up to 9 convolutional neural network (CNN) layers [75] followed by 2 fully connected (FC) layers [76]. Batch normalization [77] and weight dropout [78] were applied after all layers and max-pooling [79] after CNN layers (see tested parameters in Table S1-6). The Adam optimizer [80] with mean squared error loss function and ReLU activation function [81] with uniform [28] weight initialization were used. In total, 26 hyper-parameters were optimized over a predefined parameter space (Table S3) using a tree-structured Parzen estimators approach[82] at default settings for 1000 iterations [44,83]. The Keras v2.2 and Tensorflow v1.10 software packages were used.

## 4.9 Prediction of $T_{opt}$ for enzymes from BRENDA and CAZy

Sequences and associated OGT values for the BRENDA database was obtained from Li et al [10]. For the CAZy database, enzyme information including protein name, EC number, Organism, GenBank id, Uniprot id, PDB id and CAZy family id were obtained from http://www.cazy.org/ [45]. 1,346,471 proteins with unique GenBank identifiers were obtained. Protein sequences were firstly downloaded from NCBI ftp site: https://ftp.ncbi.nih.gov/ncbi-asn1/protein_fasta/. Then only those sequences that were present in the CAZy dataset were kept by matching GenBank identifier. 1,346,358 sequences with unique GenBank identifiers were obtained. 924,642 sequences could be mapped to an optimal growth temperature (OGT) value by cross-referencing the source organism name and an OGT dataset [66]. Only the species names were checked, ignoring strain designations, for instance *Saccharomyces cerevisiae* S288C was considered as *Saccharomyces cerevisiae*. For $T_{opt}$ prediction on the BRENDA and CAZy data the model with the best performance was selected, which in this case was the random forest model trained only on amino acid frequencies and OGT. The model was then trained on all samples in the training dataset. For the prediction, (1) the 20 amino acid frequencies were extracted with iFeature [42] and OGT values of their source organisms were mapped; (2) all those 21 features were normalized by subtracting the mean and then divided by the standard deviation obtained from the training dataset; (3) those data were used as input of the model for the prediction of the $T_{opt}$ values.

## Code and data availability

The tome package is available on GitHub (https://github.com/EngqvistLab/Tome/). The annotated $T_{opt}$ values and source organism OGTs for enzymes in the BRENDA and CAZy databases are available as flatfiles on Zenodo (https://zenodo.org/record/3578468#.XffgbpP0nOQ, DOI: 10.5281/zenodo.3578467). Other scripts and datasets are available on GitHub (https://github.com/EngqvistLab/Supplemetenary_scripts_datasets_R2LG).

## Author Contributions

GL and MKME conceptualized the research. GL and JG mathematically derived the $\langle R^2 \rangle_{LG}$. GL performed Monte Carlo simulations. GL, JZ, JL and MKME analyzed and interpreted the results of predicting enzyme $T_{opt}$. JZ and AZ analyzed and interpreted the results of transcriptomics data. GL, BJ and JN analyzed or interpreted the genomics results. GL, JZ and MKME wrote the initial draft of the paper. GL, JZ, AZ, JN and MKME carried out revisions on the initial draft and wrote the final version.

## Acknowledgements

GL and JN have received funding from the European Union's Horizon 2020 research and innovation program under the Marie Skłodowska-Curie program, project PAcMEN (grant agreement No 722287). JN also acknowledges funding from the Novo Nordisk Foundation (grant no. NNF10CC1016517), the Knut and Alice Wallenberg Foundation. JZ and AZ are supported by SciLifeLab funding. The computations were performed on resources at

# Supplementary information for:

# Performance of regression models as a function of experiment noise

Gang Li[1], Jan Zrimec[1], Boyang Ji[1], Jun Geng[1], Johan Larsbrink[1], Aleksej Zelezniak[1,2], Jens Nielsen[1,3,4], and Martin KM Engqvist[1*]

[1] Department of Biology and Biological Engineering, Chalmers University of Technology, SE-412 96 Gothenburg, Sweden

[2] Science for Life Laboratory, Tomtebodavägen 23a, SE-171 65, Stockholm, Sweden

[3] Novo Nordisk Foundation Center for Biosustainability, Technical University of Denmark, DK-2800 Kgs. Lyngby, Denmark

[4] BioInnovation Institute, Ole Måløes Vej 3, DK-2200 Copenhagen N, Denmark

* Corresponding author

E-mail: martin.engqvist@chalmers.se

## Table of contents

## Supplementary Notes

**Note 1: The expectation and variance of best $R^2$ score**

Given a set of samples with experimentally determined labels $\{y_{obs,i}\}$ and corresponding unknown real labels $\{y_i\}$, By assuming a normally distributed experimental noise term $\varepsilon_{y,i} \sim N(0, \sigma_{y,i})$, $y_{obs,i} = y_i + \varepsilon_{y,i}$ ($y_i \in R$). A complete set of features is known as $x_i \in R^k$ for each sample. The "complete" means that this set of features are sufficient to accurately calculate the real value of label $y_i$ with $y = f(x)$ for all samples. The performance of this real function $f(x)$ on the dataset $\{x_i, y_{obs,i}\}$ gives an upper bound for the expected performance of any ML model. The coefficient of determination ($R^2$) of the model $f(x)$ in the above argument is given by

$$R^2 = 1 - \frac{\sum_{i=1}^{m}(y_{obs,i} - \widehat{y}_{obs,i})^2}{\sum_{i=1}^{m}(y_{obs,i} - \bar{y}_{obs})^2} = 1 - \frac{\sum_{i=1}^{m}(y_{obs,i} - f(x_i))^2}{\sum_{i=1}^{m}(y_{obs,i} - \bar{y}_{obs})^2}$$

where $m$ is the number of samples.

$$R^2 = 1 - \frac{\sum_{i=1}^{m}(y_{obs,i} - \widehat{y}_{obs,i})^2}{\sum_{i=1}^{m}(y_{obs,i} - \bar{y}_{obs})^2}$$

Since $f(x_i) = y_i$, $y_{obs,i} - f(x_i) = y_{obs,i} - y_i = \varepsilon_{y,i}$, thereby the numerator is $\sum_{i=1}^{m}(y_{obs,i} - f(x_i))^2 = \sum_{i=1}^{m}\varepsilon_{y,i}^2$. The expectation is given by

$$\langle R^2 \rangle = 1 - \langle \frac{\sum_{i=1}^{m}\varepsilon_{y,i}^2}{\sum_{i=1}^{m}(y_{obs,i} - \bar{y}_{obs})^2} \rangle = 1 - \sum_{i=1}^{m} \langle \frac{\varepsilon_{y,i}^2}{\sum_{j=1}^{m}(y_{obs,i} - \bar{y}_{obs})^2} \rangle.$$

Since $\varepsilon_{y,i}$ is normally distributed with a zero-mean and variance of $\sigma_{y,i}^2$, then $\frac{\varepsilon_{y,i}}{\sigma_{y,i}}$ follows a standard normal distribution. Thereby $(\frac{\varepsilon_{y,i}}{\sigma_{y,i}})^2$ follows a chi-squared distribution with a degree of 1 ($\chi^2(1)$). The numerator becomes $\varepsilon_{y,i}^2 = \sigma_{y,i}^2 \frac{\varepsilon_{y,i}^2}{\sigma_{y,i}^2} \sim \sigma_{y,i}^2 \cdot \chi^2(1)$. We assume that the variance of the observed values $y_{obs,i}$ is normally distributed with a variance of $\sigma_{obs}^2$, then

$$\sum_{j=1}^{m}(y_{obs,i} - \bar{y}_{obs})^2 \sim \sigma_{obs}^2 \cdot \chi^2(m-1).$$

The ratio between two chi-squared distributions is an $F$ distribution multiplied by the ratio between their degrees of freedom, thereby

$$\langle R^2 \rangle = 1 - \sum_{i=1}^{m} \frac{\sigma_{y,i}^2}{\sigma_{obs}^2} \langle \frac{\chi^2(1)}{\chi^2(m-1)} \rangle = 1 - \sum_{i=1}^{m} \frac{\sigma_{y,i}^2}{\sigma_{obs}^2} \frac{1}{m-1} \langle F(1, m-1) \rangle.$$

Since $\langle F(1, m-1) \rangle = \frac{m-1}{m-3}$, then

$$\langle R^2 \rangle = 1 - \frac{1}{m-3} \sum_{i=1}^{m} \frac{\sigma_{y,i}^2}{\sigma_{obs}^2} = 1 - \frac{m}{m-3} \frac{\overline{\sigma_y^2}}{\sigma_{obs}^2}$$

The variance is given by

$$Var(R^2) = Var(1 - \frac{\sum_{i=1}^{m}(y_{obs,i} - f(x_i))^2}{\sum_{i=1}^{m}(y_{obs,i} - \bar{y}_{obs})^2}) = Var(\frac{\sum_{i=1}^{m}(y_{obs,i} - f(x_i))^2}{\sum_{i=1}^{m}(y_{obs,i} - \bar{y}_{obs})^2})$$

With a similar approach as for expectation,

$$Var(R^2) = Var(\sum_{i=1}^{m} \frac{\sigma_{y,i}^2}{\sigma_{obs}^2} \frac{1}{m-1} \frac{\varepsilon_{y,i}^2/\sigma_{y,i}^2}{(\sum_{i=1}^{m}(y_{obs,i} - \bar{y}_{obs})^2/\sigma_{obs}^2)/(m-1)})$$

$\frac{\sigma_{y,i}^2}{\sigma_{obs}^2}$ is a constant and the expectation of the ratio part is $\frac{2(m-1)^2(m-2)}{(m-3)^2(m-5)}$, thereby

$$Var(R^2) = \frac{1}{(m-1)^2} \frac{2(m-1)^2(m-2)}{(m-3)^2(m-5)} \sum_{i=1}^{m} \frac{\sigma_{y,i}^4}{\sigma_{obs}^4} = \frac{2m(m-2)}{(m-3)^2(m-5)} \frac{\overline{\sigma_y^4}}{\sigma_{obs}^4}$$

**Note 2: The expectation and variance of MSE**

The expectation of MSE on the dataset $\{x_i, y_{obs,i}\}$ is given by

$$\langle MSE \rangle = \langle \frac{1}{m} \sum_{i=1}^{m} (y_{obs,i} - f(x_i))^2 \rangle$$

Since $f(x_i) = y_i$, $y_{obs,i} - f(x_i) = y_{obs,i} - y_i = \varepsilon_{y,i}$, thereby

$$\langle MSE \rangle = \frac{1}{m} \langle \sum_{i=1}^{m} \varepsilon_{y,i}^2 \rangle = \frac{1}{m} \sum_{i=1}^{m} \sigma_{y,i}^2 \langle \frac{\varepsilon_{y,i}^2}{\sigma_{y,i}^2} \rangle$$

Since $\varepsilon_{y,i}$ is normally distributed with a zero mean and variance of $\sigma_{y,i}^2$, then $\frac{\varepsilon_{y,i}}{\sigma_{y,i}}$ follows a standard normal distribution. Thereby $(\frac{\varepsilon_{y,i}}{\sigma_{y,i}})^2$ follows a chi-squared distribution with a degree of 1 ($\chi^2(1)$). The expectation of this $\chi^2(1)$ is 1, thereby

$$\langle MSE \rangle = \frac{1}{m} \sum_{i=1}^{m} \sigma_{y,i}^2 = \overline{\sigma_y^2}$$

This gives a lower bound of expected MSE values for machine learning models.

Accordingly the variance of MSE is given by

$$Var(MSE) = Var(\frac{1}{m} \sum_{i=1}^{m} (y_{obs,i} - f(x_i))^2) = Var(\frac{1}{m} \sum_{i=1}^{m} \sigma_{y,i}^2 \frac{\varepsilon_{y,i}^2}{\sigma_{y,i}^2})$$

The $\sigma_{y,i}^2$ is a constant and the variance of $\frac{\varepsilon_{y,i}^2}{\sigma_{y,i}^2} \sim \chi^2(1)$ is 2. Thereby

$$Var(MSE) = \frac{2}{m^2} \sum_{i=1}^{m} \sigma_{y,i}^4 = \frac{2\overline{\sigma_y^4}}{m}$$

## Supplementary Figures

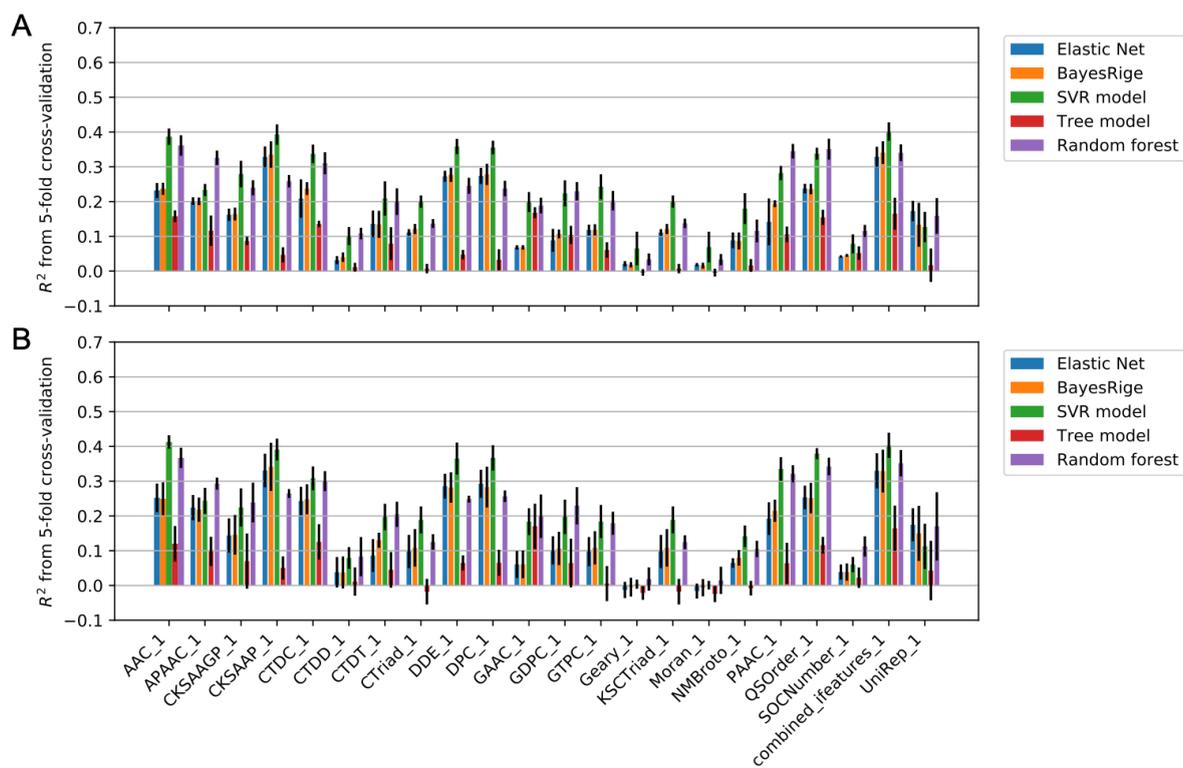

**Figure S1.** The performance five regression models when trained on different feature sets **without OGT** as an additional feature. (A) the dataset before cleaning; (B) the dataset after data cleaning.

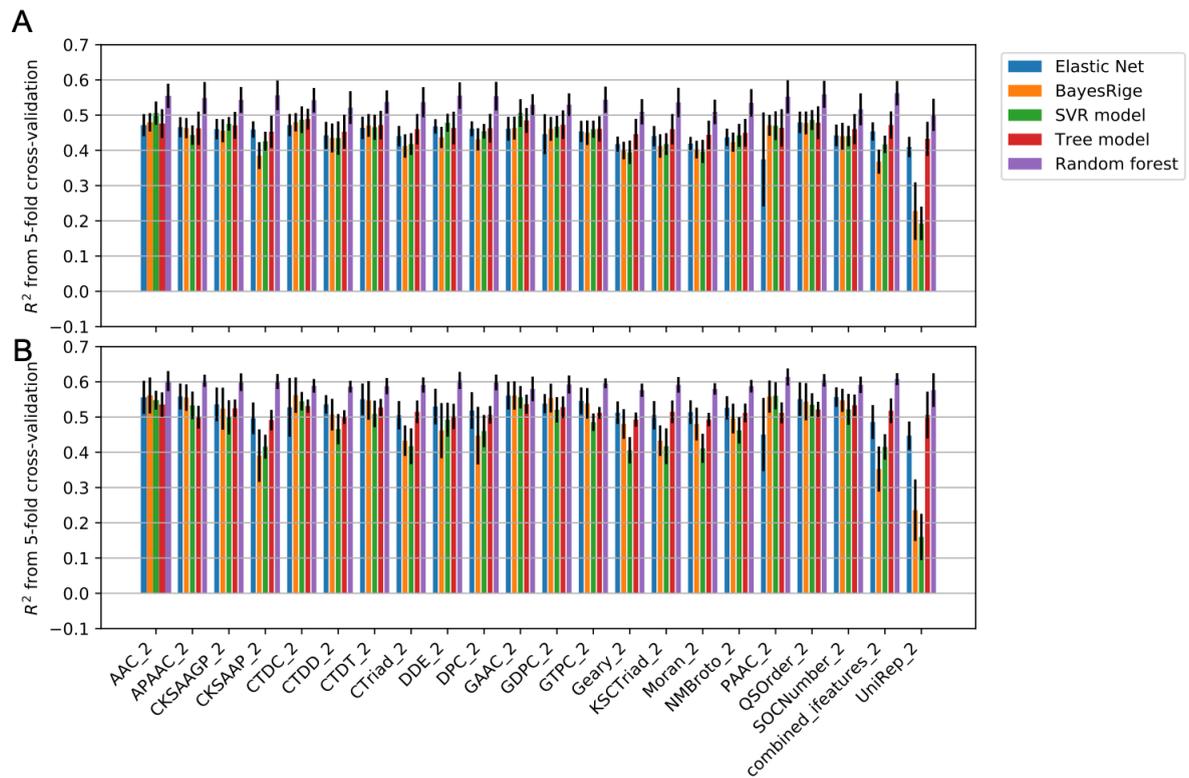

**Figure S2.** The performance five regression models when trained on different feature sets **with OGT** as an additional feature. (A) the dataset before cleaning; (B) the dataset after data cleaning.

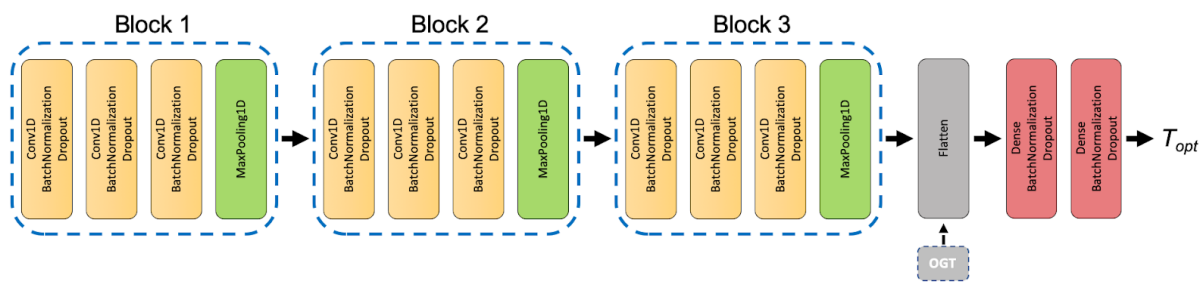

Figure S3. Deep NN architecture. There are three convolution layers in each of three blocks and have the same hyper-parameters. The hyper-parameter space for optimization with Hyperopt[1] is listed in Table S3.

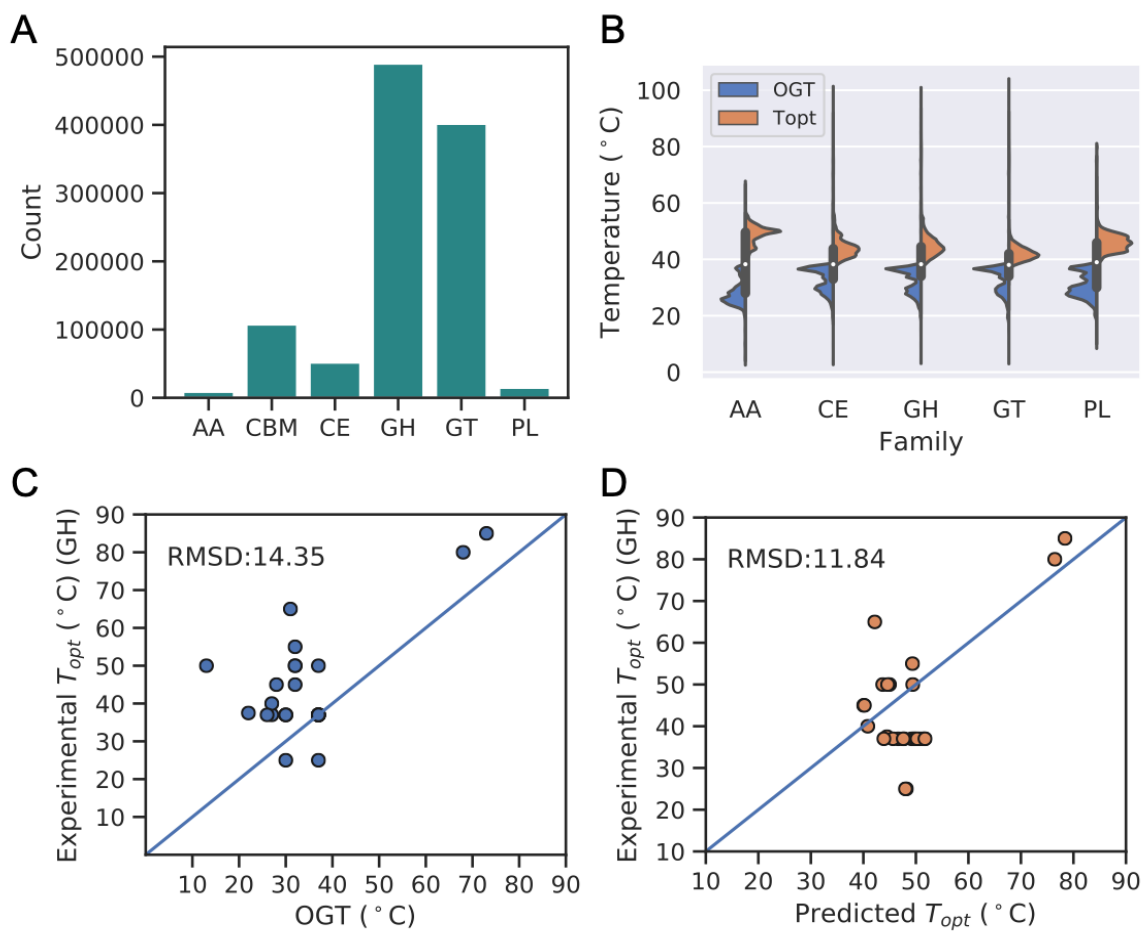

**Figure S4.** Predict $T_{opt}$ of CAZy enzymes [2]. (A) 924, 642 sequences covering 6 CAZy families can be mapped to an optimal growth temperature (OGT) value by cross-referencing the source organism name and an OGT dataset[3]. The distribution of OGT and predicted $T_{opt}$ of each CAZy family was shown in (B). A list of commercialized enzyme $T_{opt}$ values from nzytech (https://www.nzytech.com/) were collected to validate our predictions. nzytech data were downloaded from https://www.nzytech.com/resources/catalogues/. A pdf file cazymes_2019.pdf was downloaded. Then this pdf file was parsed to obtain the CAZy family id, source organism name and optimal temperature of all enzymes in the file. Since there is no sequence provided, nor any sequence/gene id that could be mapped to a sequence database, it's impossible to exactly map those enzymes to the ones in CAZy database. Thereby we used the following strategies to do the mapping: for a given CAZy family id from a specific organism, if there is only one record in nzytech dataset and also only one record in CAZy dataset, then we consider those two enzymes are the same enzyme. In such a way, we could find experimental $T_{opt}$ values from nzytech dataset. To validate our prediction, the enzymes in the training dataset were also removed by comparing protein sequences of those CZAy enzymes to ones in the training dataset. In the end, 27 enzymes from family GH were obtained (there are only less than 10 enzymes were found for other families, then they are not included in comparison) . Even though our prediction is still not a perfect estimation of experimental values ( RMSE: 11.84 °C ) , this is a more accurate estimation than OGT values (Figure S3C and S3D). AA: Auxiliary Activity, CBM: Carbohydrate-Binding Module, CE: Carbohydrate Esterase, GH: Glycoside Hydrolase, GT: Glycosyl Transferase, PL: Polysaccharide Lyase.

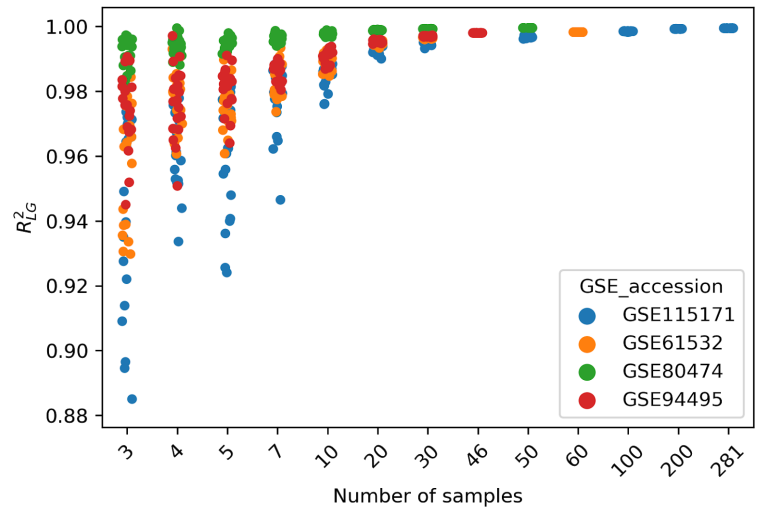

**Figure S5.** The estimated upper bounds for condition-specific subsets of the transcriptomics dataset. NCBI Gene Expression Omnibus (https://www.ncbi.nlm.nih.gov/geo/) identifiers GSE were used to group the data across conditions and $\langle R^2 \rangle_{LG}$ was estimated for the four largest subsets with 281, 60, 50 and 46 samples, respectively.

Li et al. 2019 - Supplementary Information

## Supplementary Tables

**Table S1.** The estimated $\langle R^2 \rangle_{LG}$ for melting temperature datasets from Leuenberger *et al* [4].

|  | $\overline{\sigma_y^2}$ | $\sigma_{obs}^2$ | $\langle R^2 \rangle_{LG}$ |
|---|---|---|---|
| *S. cerevisiae* | $1.56^2$ | $5.89^2$ | 0.93 |
| *E. coli* | $1.31^2$ | $7.39^2$ | 0.97 |
| Human Hela cell | $5.49^2$ | $6.57^2$ | 0.30 |
| *T. thermophilus* | $1.29^2$ | $8.02^2$ | 0.97 |

**Note:** Leuenberger R et al [4] measured melting temperatures ($T_m$ values) of 3,557 proteins from *Escherichia coli* (730), *Saccharomyces cerevisiae* (707), *Thermus thermophilus* (1,083), and human Hela cells (1,037) via a proteomics approach. In this approach, proteins were first digested into peptides by limited proteolysis. Then $T_m$s of those peptides were measured. Thirdly, peptides with high-quality $T_m$ values were clustered the average $T_m$ were assigned as the $T_m$ of this cluster. At last, the cluster with the lowest $T_m$ was assigned as the $T_m$ of the protein. Since the standard error was not reported for protein $T_m$ values, the reported 95% confidence interval of single peptides were used to estimate the standard error of protein $T_m$ values with following approach: 1) calculate the standard error of each peptide listed Table S3 of [4] from its 95% confidence interval listed in Table S3 of [4] as (tm_ciu-tm_cil)/2/1.96, in which tm_ciu and tm_cil are the upper and lower bounds; 2) for a dataset with a list of proteins from considered organism(s), calculate the average squared standard errors of the peptides in the dataset; 3) estimate the average number of peptides in each protein in the considered dataset by dividing the number of peptides in each protein by the theoretical number of domains (from Table S3 of [4]); 4) the average peptide standard error from step 2) was divided by the root of the average peptide number obtained from step 3). This value was considered as an approximation of the average standard error $\sqrt{\overline{\sigma_y^2}}$ of the considered dataset.
10

**Table S2.** Regression models and corresponding hyper-parameter space searched.

| Regression model | Module | Hyperparameter range |
| --- | --- | --- |
| Linear model | sklearn.linear_model.LinearRegression | None |
| Elastic net | sklearn.linear_model.ElasticNetCV | Default |
| Bayes ridge | sklearn.linear_model.BayesianRidge | None |
| Support vector regressor | sklearn.svm.SVR | 'C': numpy.logspace(-5, 10, num=16, base=2.0), 'Epsilon': [0, 0.01, 0.1, 0.5, 1.0, 2.0, 4.0] |
| Decision tree | sklearn.tree.DecisionTreeRegressor | 'Min_samples_leaf': numpy.linspace(0.01, 0.5, 10) |
| Random forest | sklearn.ensemble.RandomForestRegressor | 'Max_features': numpy.arange(0.1, 1.1, 0.1) |

**Table S3.** Searched hyper-parameter space of deep neural network (Figure S3)

|  | parameter | Range |
|---|---|---|
| Block 1 | kernel size | [20, 30, 40] |
|  | filter | [32, 64] |
|  | stride | [2, 4, 8] |
|  | dilation | [1, 2, 4] |
|  | pool size | [2, 4, 8] |
|  | drop out | (0, 0.4) |
| Block 2 | kernel size | [10, 20, 30] |
|  | filter | [64, 128] |
|  | stride | [1, 2] |
|  | dilation | [1, 2, 4] |
|  | pool size | [1, 2, 4] |
|  | drop out | (0, 0.4) |
| Block 3 | kernel size | [10, 20] |
|  | filter | [128, 256] |
|  | stride | [1, 2] |
|  | dilation | [1, 2, 4] |
|  | pool size | [1, 2, 4] |
|  | drop out | (0, 0.4) |
| 1st dense layer | size | [64, 128] |
|  | drop out | (0, 0.3) |
| 2nd dense layer | size | [32, 64] |
|  | drop out | (0, 0.3) |

## Supplementary References